# Low-Data OOD Prediction of SAW Resonators with Simulation-Compensated Transfer Learning

Zhijie Ma[1], Xingyu Liu[1,*], Shuhan Chen[1], Fangsheng Qian[1], Junyan Zheng[1], Nan Jun Kim[2], Taek Joo Lee[2], Hyung Kook Yang[2] and Yansong Yang[1,*]

[1]*Department of Electronic and Computer Engineering, The Hong Kong University of Science and Technology, Hong Kong, China*

[2]*SAWNICS Inc., Korea*

Email: {zmaal, xliufs, shuhan.chen}@connect.ust.hk, {eefqian, eejzheng}@ust.hk, {kimnj, tjlee, yanghk}@sawnics.com, eeyyang@ust.hk

[*]Corresponding authors: Xingyu Liu and Yansong Yang.

***Abstract*—Surface acoustic wave (SAW) resonators are foundational to modern radio-frequency (RF) front-end architectures. However, models trained on limited fabricated devices generalize poorly to unmeasured design regions because of process-induced variations, while finite-element method (FEM) simulations, although inexpensive, exhibit systematic offsets from measurements. To predict device performance before physical prototyping, this work presents a simulation-compensated transfer-learning framework that uses computationally efficient 2.5-D FEM simulations as physics-based anchors and learns their calibration to the measurement domain. Using 4,236 measured lithium tantalate ($LiTaO_3$) SAW resonators and 816 unique simulated configurations, five transfer strategies are systematically benchmarked under seven out-of-distribution (OOD) protocols spanning frequency, geometry, and material shifts. The proposed framework reduces OOD prediction error by 83.54% on average (up to 97.57%) relative to the Measurement-only baseline, with the best strategy selected per case, and raises the pooled $R^2$ of frequency prediction to 1.000, offering an economically scalable route toward physics-guided, fabrication-aware machine learning for microelectromechanical-system (MEMS) acoustic-device design.**

## I. INTRODUCTION

Surface acoustic wave (SAW) resonators are indispensable components of modern wireless front ends, owing to their compact footprint, strong electromechanical coupling, and compatibility with wafer-scale semiconductor manufacturing. They underpin radio-frequency (RF) filters, duplexers, and oscillators throughout current 5G systems and are expected to remain critical for 6G front-end architectures spanning increasingly diverse frequency bands. Continued advances in piezoelectric substrates, electrode configurations, and wafer-level processing have further improved achievable quality factor, coupling coefficient, and power handling, enabling ever more stringent RF specifications to be met [1]–[3].

Resonator development nonetheless relies heavily on iterative fabrication and measurement. Device performance is jointly determined by structural parameters, material properties, and process variations introduced during lithography, deposition, and etching, causing resonators with nearly identical layouts routinely to exhibit measurable deviations in resonance frequency, anti-resonance frequency, and electromechanical coupling after fabrication. Capturing these process-dependent effects requires fabrication and measurement across multiple fabrication iterations, constrained by slow and costly experimental data acquisition [4], [5]. Regression models, deep networks, and physics-aware surrogates predict resonator characteristics accurately once sufficient labeled data become available, as demonstrated in a companion large-scale study for lithium tantalate ($LiTaO_3$) SAW resonators [6].

Nearly all such approaches, however, implicitly assume that training and testing samples are independently and identically distributed (IID), which is an assumption rarely satisfied when engineers routinely extrapolate to new operating frequencies, untested geometries, or different piezoelectric cuts in practice. As illustrated in Fig. 1, out-of-distribution (OOD) evaluation exposes a severe weakness common to standard regression families: whereas linear, random-forest, and gradient-boosted models all reproduce measured performance under IID conditions, their normalized mean absolute error (MAE) inflates by one to three orders of magnitude relative to the IID baseline MAE once test samples fall outside the geometry, material, or frequency range represented in training. This vulnerability substantially limits deploying conventional machine learning in industrial resonator development.

Mitigating this OOD vulnerability requires either expanding measured coverage or supplying additional physical information beyond what limited fabrication data can provide. Expanding measured coverage by fabricating additional wafers to exhaustively cover every structural combination and material platform incurs prohibitive cost and long development cycles.

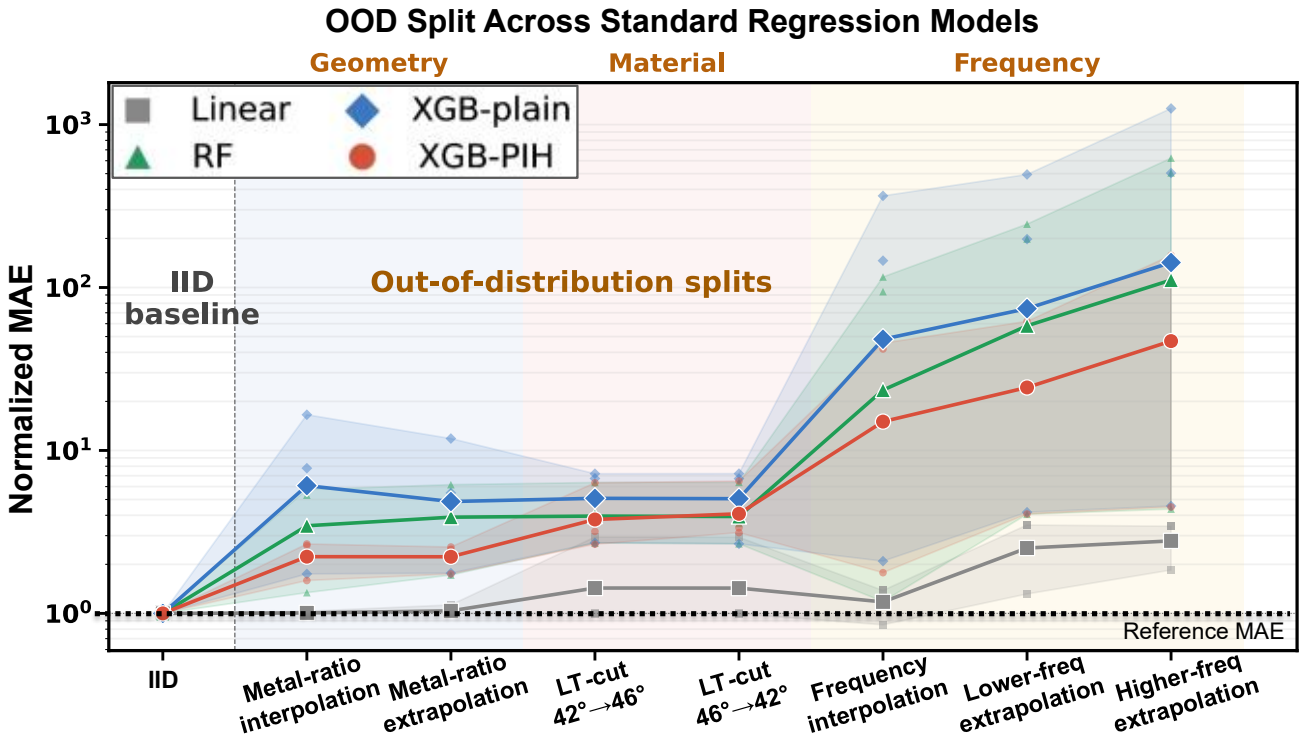

Fig. 1. Normalized MAE of OOD splits across all standard regression families: linear, random forest, and plain (XGB-plain) and physics-informed-feature (XGB-PIH) extreme gradient boosting (XGBoost).

Finite-element method (FEM) simulation offers an attractive complementary source by providing inexpensive, noise-free, and densely sampled structural data while preserving the physical relationships governing acoustic-wave propagation [7]. Simulated responses alone nevertheless diverge from measured device characteristics due to simplified material models, idealized boundary conditions, and neglected process imperfections. Therefore, directly pooling simulated and measured samples typically yields only limited benefit and may even degrade performance.

Rather than treating simulation and measurement as interchangeable datasets, this work treats simulation as a structured source-domain prior that supplies transferable physical knowledge for measurement-domain learning. Simulated responses cannot reproduce fabricated-device characteristics in absolute terms, yet they faithfully preserve the physical trends relating geometry to performance. Transferring these trends through appropriate calibration is hypothesized to regularize machine learning models and to improve generalization beyond the measured design space.

The remainder of this paper is organized as follows. Section II describes the Measurement-only baseline and simulation-compensated transfer-learning framework. Section III presents results across seven OOD protocols, and Section IV concludes the paper.

## II. SIMULATION-COMPENSATED TRANSFER LEARNING FRAMEWORK

As shown in Fig. 2, resonator performance prediction can follow either a conventional Measurement-only loop or the proposed simulation-compensated loop. In the conventional workflow, a regression model is trained exclusively on limited fabricated data and queried directly for OOD prediction, a path that yields degraded generalization, as quantified in Fig. 1. The simulation-compensated workflow instead couples the regressor with inexpensive data-efficient FEM simulation, embedding physical understanding into the learning process before OOD prediction is attempted. This section first reviews the conventional flow and its governing relations (Section II-A), and then presents the proposed simulation-compensated workflow (Section II-B).

### *A. Conventional Measurement-only Prediction Flow*

The conventional flow begins from an experimental database collected through wafer-scale fabrication of $LiTaO_3$ SAW resonators [6]. After removing incomplete and invalid records, 4,236 resonators are retained, each combining structural parameters including electrode pitch, metallization ratio, aluminum electrode thickness, and number of interdigital transducer (IDT) pairs, across both 42° and 46° YX lithium tantalate cuts, with electrical characteristics extracted from vector network analyzer (VNA) measurements. The Measurement-only baseline is trained directly on the measured database to map structural parameters onto the resonance frequency $f_s$, the anti-resonance frequency $f_p$, and the electromechanical coupling coefficient $k_t^2$, following the conventional design pattern of learning the complete measurement-domain mapping from fabricated samples alone.

Although the conventional flow performs adequately under IID evaluation, it suffers a severe MAE inflation when scaled to the OOD regime, as quantified in Fig. 1. This vulnerability reflects a structural limitation rather than a deficiency of any specific model, motivating a second regime that couples limited fabrication data with inexpensive simulation to recover robust generalization without additional fabrication cost.

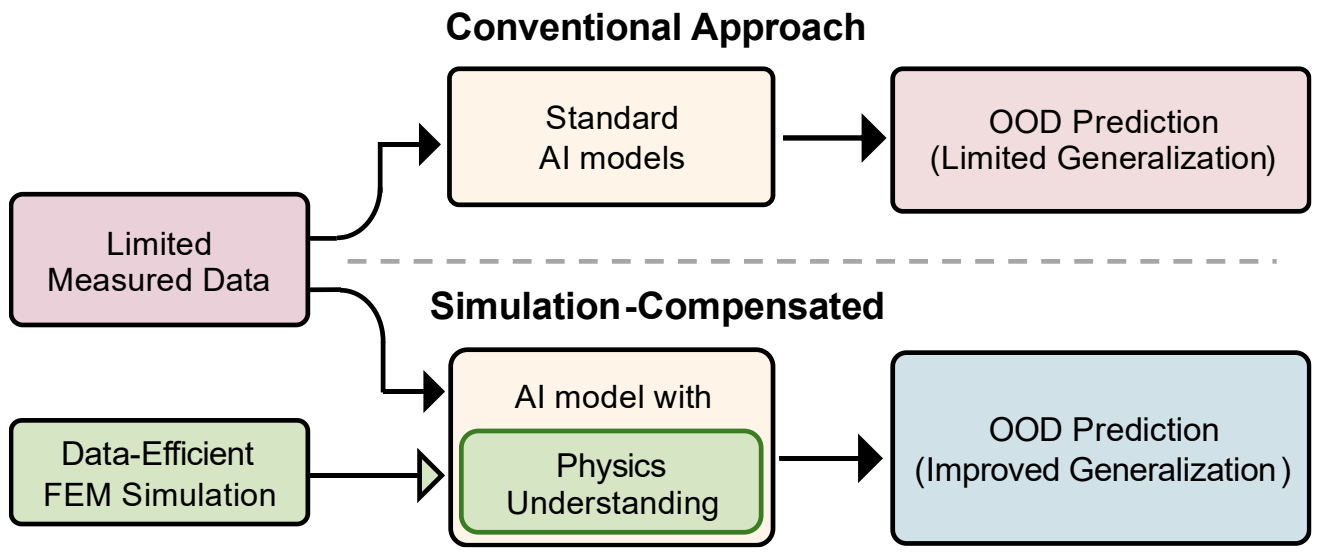

Fig. 2. Conventional Measurement-only AI loop versus the proposed simulation-compensated loop.

### *B. Simulation-Compensated Transfer Learning Workflow*

Addressing the preceding limitation requires a source of physical information that extends beyond the measured design space at negligible additional cost. A physics-based simulation database is accordingly established using 2.5-D periodic finite-element analysis in COMSOL Multiphysics, featuring one IDT pair with periodic boundary conditions and a perfectly matched layer (PML) at the substrate bottom. This model substantially reduces computational cost relative to full three-dimensional device simulation while accurately capturing the dominant acoustic behavior of an infinite IDT structure. In total, 816 simulated configurations are generated by sweeping the four structural variables shared with the measured database: substrate cut, pitch, metallization ratio, and aluminum thickness. The number of IDT pairs is not swept for computational efficiency. Although idealized material properties and simplified boundary conditions introduce a persistent domain gap, and the simulation exhibits systematic offsets relative to measurement, the simulation database preserves the intrinsic dependence of resonator performance on structural parameters, and is therefore exploited not as a source of absolute performance values, but as a source-domain representation of transferable physical trends.

The framework treats simulation and measurement as two related, statistically distinct domains. Let x denote the structural parameter vector (pitch, metallization ratio, aluminum thickness, number of IDT pairs, substrate cut), $y_{\mathrm{m}}$ the measured response, and $y_s$ the simulation prediction. Rather than estimating $y_{\mathrm{m}} = f(x)$ directly from limited measured data, the framework learns a family of transfer functions $y_m = T(x, y_s)$. Let $\varphi(x)$ denote the physics-informed measurement features, s the simulated-response vector, $s_\gamma$ the simulated response for the current target, and $z = [\varphi(x), s]$.

Direct fusion augments the measurement features with simulated responses and is expressed in (1).

$$\hat{y}_d = f_d(z) \tag{1}$$

Residual learning models the additive simulation-to-measurement discrepancy, as defined in (2).

$$\Delta = y_m - s_y, \quad \hat{y}_\Delta = s_y + f_\Delta(\mathrm{z}) \tag{2}$$

Ratio-based calibration models the corresponding multiplicative discrepancy, as defined in (3), where $\varepsilon = 10^{-2}$ prevents division by zero.

$$r = \frac{y_m}{\max(s_y, \varepsilon)}, \quad \hat{y}_r = s_y\, f_r(\mathrm{z}) \tag{3}$$

Mean-anchor correction in (4) recenters the predicted residual distribution; $\mu_{tr}$ and $\mu_{te}$ denote sample means over the training and test batches, respectively.

$$\hat{y}_{a,i} = s_{y,i} + \Delta_i - \mu_{te}(\hat{\Delta}) + \mu_{tr}(\Delta) \tag{4}$$

Weighted fusion in (5) combines simulation-assisted predictors using inverse-squared validation-MAE weights, where $e_j$ is the validation MAE of method j.

$$w_j = \frac{e_j^{-2}}{\Sigma_k e_k^{-2}}, \quad \hat{y}_w = \Sigma_j w_j\, \hat{y}_j, \quad e_j = \mathrm{MAE}_{\mathrm{val},j} \tag{5}$$

Rather than presupposing a single universal mechanism, this work benchmarks the resulting method bank under the evaluation protocol in Table I and retrospectively reports the best-performing method for each target–protocol combination as an oracle envelope.

TABLE I
SEVEN-SPLIT OUT-OF-DISTRIBUTION EVALUATION PROTOCOL

| Split | Description | OOD Type | Train / Test |
|---|---|---|---|
| (i) | High-frequency extrapolation | Frequency | 3,516 / 720 |
| (ii) | Low-frequency extrapolation | Frequency | 3,444 / 792 |
| (iii) | Frequency interpolation | Frequency | 3,449 / 787 |
| (iv) | Metallization-ratio interpolation | Geometry | 2,729 / 1,507 |
| (v) | Metallization-ratio boundary extrapolation | Geometry | 2,113 / 2,123 |
| (vi) | 42°→46° YX $LiTaO_3$ transfer | Material | 2,117 / 2,119 |
| (vii) | 46°→42° YX $LiTaO_3$ transfer | Material | 2,119 / 2,117 |

## III. EXPERIMENTAL VALIDATION AND RESULTS

The Measurement-only baseline (Section II-A) uses measured data alone, without simulation-based calibration. The five transfer strategies (Section II-B) are evaluated on the same seven splits (Table I) for $f_s$, $f_p$, and $k_t^2$, using the coefficient of determination $R^2$ and MAE, defined in (6) and (7).

$$R^2 = 1 - \frac{\Sigma_{i=1}^{N} (y_i - \hat{y}_i)^2}{\Sigma_{i=1}^{N} (y_i - \bar{y})^2} \tag{6}$$

$$\mathrm{MAE} = \frac{1}{N} \Sigma_{i=1}^{N} |y_i - \hat{y}_i| \tag{7}$$

Here, $N$ denotes the number of held-out samples, $y_i$ and $\hat{y}_i$ denote the measured and predicted responses, and $\bar{y}$ is the mean measured response. Pooled metrics are computed after concatenating held-out predictions from all seven OOD protocols. The cross-target summaries are arithmetic means of the three target-specific pooled $R^2$ values and relative MAE reductions, respectively; raw MAEs are not averaged because their units differ (MHz for $f_s$ and $f_p$ and percentage points for $k_t^2$).

Fig. 3 compares measured-vs-predicted scatter for all seven OOD splits pooled together, contrasting the Measurement-only baseline (top row) against the best per-cell simulation-compensated strategy (bottom row). A consistent reduction in prediction error is observed for all three targets, with the mean pooled $R^2$ increasing from 0.725 to 0.934. Fig. 4 reports the pooled percentage MAE reduction achieved by each transfer strategy relative to Measurement-only. For $f_s$ and $f_p$, every strategy except direct fusion cuts pooled MAE by more than 92%, attained without acquiring any additional fabrication data. Direct fusion, defined in (1), regresses measurements directly onto raw simulated values, and achieves only less than 20% MAE reduction. This gap indicates that the network exploits the physical trend represented by the simulation, not its absolute magnitude. The narrow gaps among the other four strategies further suggest that the simulation-compensated improvement is largely structural and systematic, such that pooled MAE reduction is not strongly sensitive to the specific calibration algorithm once the model has access to the physics trend. Coupling-coefficient prediction follows the same ranking among strategies but with markedly smaller reductions, since the MAE is already small from the baseline prediction, leaving less room for improvement.

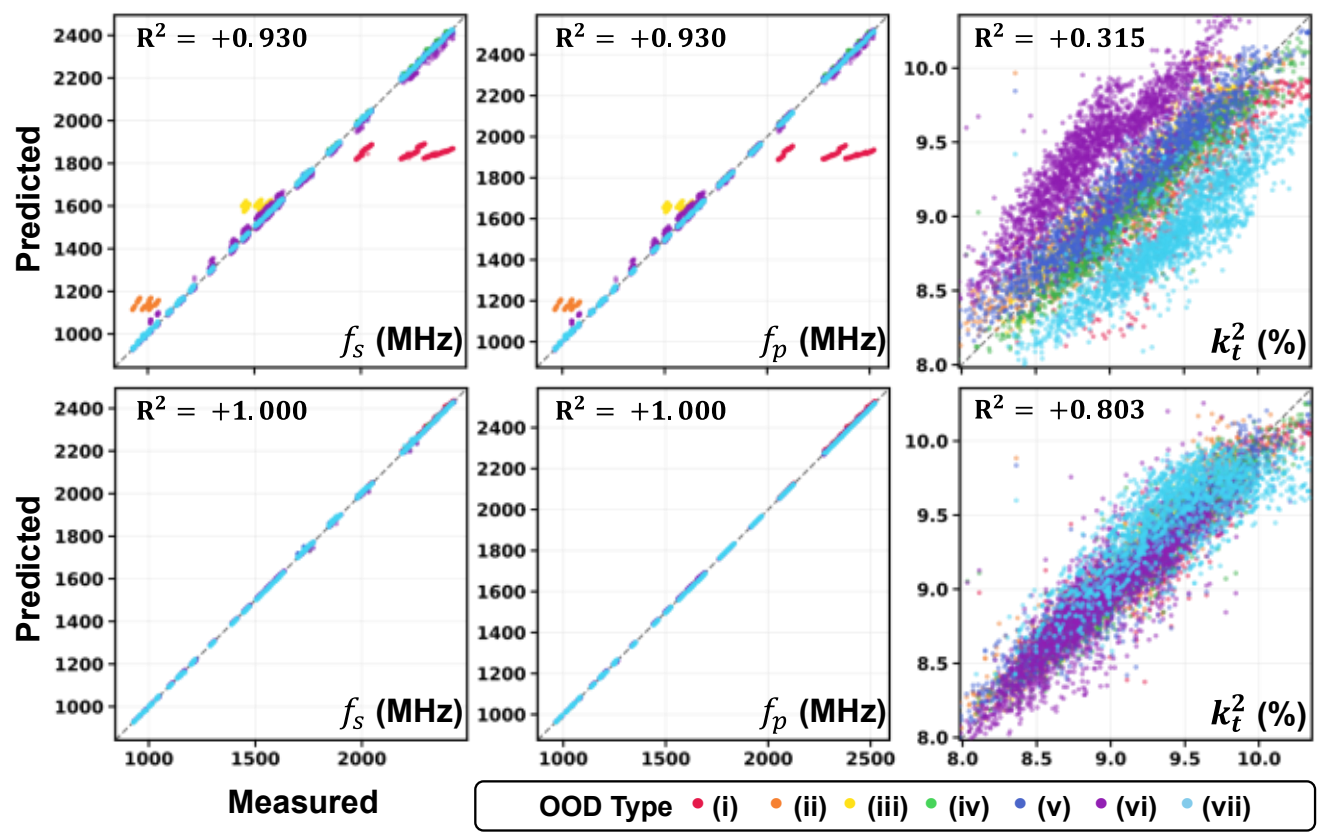

Fig. 3. Measured-vs-predicted comparison pooled across all seven OOD splits and three targets: Measurement-only (top) vs. retrospective best per-cell simulation-compensated strategy (bottom).

Table II lists the best-performing strategy for each split and target, together with the MAE and $R^2$. Resonance and anti-resonance frequencies depend primarily on acoustic wavelength and effective phase velocity [8], and Measurement-only regressors infer these relationships from limited measured coverage, which leads to large frequency errors under OOD evaluation. The FEM model captures the dominant wavelength-dependent frequency trend despite systematic offsets arising from idealized material properties and boundary conditions. Simulation compensation therefore supplies the missing physical knowledge and lets measurement-domain learning

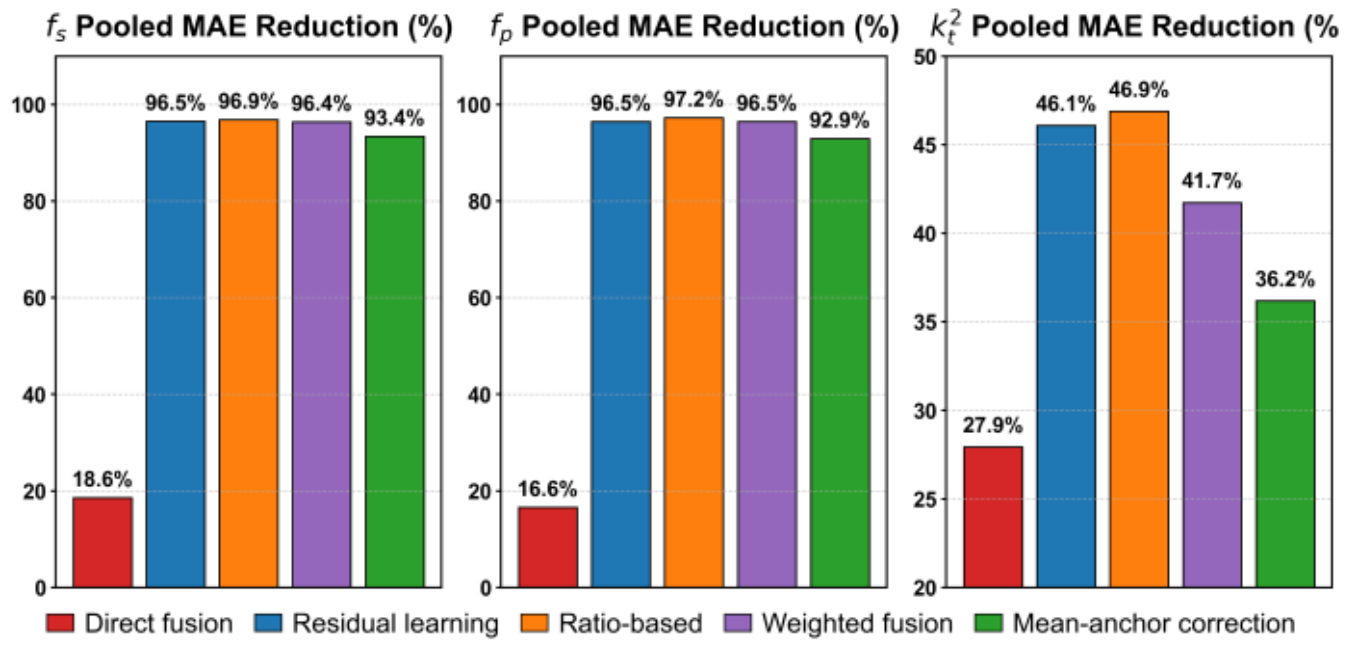

Fig. 4. Pooled percentage MAE reduction of the five evaluated transfer strategies across all OOD test splits for $f_s$, $f_p$, and $k_t^2$.

focus on calibrating the remaining discrepancy.

In contrast, $k_t^2$ largely depends on piezoelectric constants, electrode loading, and process-dependent effects [8]. When an OOD split changes the operating frequency while retaining the material platform, coupling can therefore remain comparatively stable even as absolute frequency prediction deteriorates. With smaller margin for improvement, the best-performing strategy varies across OOD protocols. Despite this, simulation compensation raises the baseline $R^2$ at the cross-material transfer in splits (vi) and (vii) from negative baseline to positive values, indicating that simulation enables the calibrated model to generalize beyond the material platform.

TABLE II
BEST-PERFORMING TRANSFER STRATEGY, MAE, AND $R^2$ PER SPLIT AND TARGET

| Split | Target | Baseline MAE | Baseline $R^2$ | Best Strategy | Best MAE | Best $R^2$ |
|---|---|---|---|---|---|---|
| (i) | $f_s$ | 360.61 | -6.4720 | Residual learning | 2.98 | 0.9992 |
| | $f_p$ | 376.10 | -6.4868 | Ratio-based | 1.89 | 0.9996 |
| | $k_t^2$ | 0.30 | 0.3367 | Weighted fusion | 0.16 | 0.6569 |
| (ii) | $f_s$ | 144.76 | -16.6758 | Ratio-based | 0.48 | 0.9997 |
| | $f_p$ | 150.27 | -16.9873 | Ratio-based | 0.40 | 0.9998 |
| | $k_t^2$ | 0.15 | 0.8573 | Ratio-based | 0.11 | 0.9077 |
| (iii) | $f_s$ | 86.60 | -3.3655 | Ratio-based | 0.56 | 0.9998 |
| | $f_p$ | 89.30 | -3.2789 | Ratio-based | 0.66 | 0.9997 |
| | $k_t^2$ | 0.11 | 0.8824 | Ratio-based | 0.09 | 0.9161 |
| (iv) | $f_s$ | 5.22 | 0.9998 | Mean-anchor | 0.78 | 1.0000 |
| | $f_p$ | 4.60 | 0.9998 | Ratio-based | 0.54 | 1.0000 |
| | $k_t^2$ | 0.13 | 0.8375 | Direct fusion | 0.10 | 0.8856 |
| (v) | $f_s$ | 3.90 | 0.9998 | Mean-anchor | 1.21 | 1.0000 |
| | $f_p$ | 3.52 | 0.9999 | Residual learning | 0.70 | 1.0000 |
| | $k_t^2$ | 0.14 | 0.7973 | Ratio-based | 0.11 | 0.8371 |
| (vi) | $f_s$ | 8.80 | 0.9987 | Residual learning | 1.67 | 1.0000 |
| | $f_p$ | 9.00 | 0.9987 | Mean-anchor | 1.77 | 1.0000 |
| | $k_t^2$ | 0.54 | -0.8806 | Weighted fusion | 0.15 | 0.6245 |
| (vii) | $f_s$ | 2.79 | 0.9999 | Residual learning | 1.69 | 1.0000 |
| | $f_p$ | 3.55 | 0.9999 | Mean-anchor | 1.85 | 1.0000 |
| | $k_t^2$ | 0.53 | -0.8801 | Residual learning | 0.20 | 0.6201 |

*Note: MAE is reported in MHz for $f_s$ and $f_p$, and in percentage points for $k_t^2$.*

Weighting each split-level MAE in Table II by its held-out sample size in Table I gives the pooled MAE attained by the best strategy at each cell. The baseline-to-best-strategy MAE falls from 47.53 to 1.36 MHz for $f_s$ (97.14% reduction), from 49.29 to 1.20 MHz for $f_p$ (97.57% reduction), and from 0.313% to 0.138% for $k_t^2$ (55.91% reduction). Averaging these three targets gives 83.54% pooled reductions. This evaluation pools the best-performing strategy selected independently for each split, whereas each bar in Fig. 4 pools a single fixed strategy applied uniformly across all seven splits. Since no single transfer mechanism dominates every scenario, this result exceeds the best fixed-strategy reduction in Fig. 4 by about three percentage points, confirming that the improvement is largely structural rather than specific to one algorithm.

These results support treating simulation and measurement as complementary representations of a single physical system, rather than as competing information sources. FEM simulation captures idealized physics-based trends, whereas measurement incorporates fabrication-dependent variation; transfer learning bridges the two, letting measurement calibrate simulation-to-fabrication deviations. As experimental data for new material platforms or process nodes become increasingly impractical to collect exhaustively, while multiphysics simulation continues to grow cheaper, simulation-compensated learning can reduce reliance on additional fabrication data, complementing existing spectrum-reconstruction, spurious-mode, and inverse-design capabilities on the same platform [6].

## IV. CONCLUSION

A simulation-compensated transfer learning framework is developed for low-data OOD prediction of lithium tantalate SAW resonators, combining the measurement database with FEM simulation and five complementary transfer strategies. The framework directly transfers physical trends learned from simulated configurations into the measurement domain, requiring no modification to existing fabrication or simulation workflows. Experimental validation on fabricated $LiTaO_3$ resonators across seven representative OOD protocols shows substantially improved prediction of resonance frequency, anti-resonance frequency, and electromechanical coupling relative to Measurement-only, with the best strategy selected retrospectively per case, increasing the mean pooled $R^2$ from 0.725 to 0.934 and reducing MAE by an average of 83.54% across the three targets. These results indicate that the principal value of finite-element simulation resides not in its absolute numerical accuracy, but in its capacity to encode transferable physical relationships that remain valid beyond the available fabrication data, supporting physics-guided, fabrication-aware machine learning for microelectromechanical-system (MEMS) acoustic-device design.

## ACKNOWLEDGEMENT

We acknowledge the funding support in part by the Hong Kong Research Grants Council under Grant C6002-25, in part by the National Natural Science Foundation of China under Grant 62304193, and in part by the National Key Research and Development Program of China under Grant 2025YFE0201100.

## REFERENCES

[1] K. Yang *et al.*, "Advanced RF filters for wireless communications," *Chip*, vol. 2, no. 4, Dec. 2023, Art. no. 100058, doi: 10.1016/j.chip.2023.100058.

[2] F. Qian *et al.*, "Suppressing acoustomigration and temperature rise for high-power robust acoustics," *Nat. Commun.*, vol. 17, Art. no. 5543, Apr. 2026, doi: 10.1038/s41467-026-72102-7.

[3] S. Gong and G. Piazza, "Design and analysis of lithium–niobate-based high electromechanical coupling RF-MEMS resonators for wideband filtering," *IEEE Trans. Microw. Theory Techn.*, vol. 61, no. 1, pp. 403–414, Jan. 2013, doi: 10.1109/TMTT.2012.2228671.

[4] A. Kochhar, A. Mahmoud, Y. Shen, N. Turumella, and G. Piazza, "X-cut lithium niobate-based shear horizontal resonators for radio frequency applications," *J. Microelectromech. Syst.*, vol. 29, no. 6, pp. 1464–1472, Dec. 2020, doi: 10.1109/JMEMS.2020.3026167.

[5] F. Sui, W. Yue, R. Guo, K. Behrouzi, and L. Lin, "Designing weakly coupled MEMS resonators with machine learning-based method," in *Proc. IEEE 35th Int. Conf. Micro Electro Mech. Syst. (MEMS)*, Jan. 2022, pp. 454–457, doi: 10.1109/MEMS51670.2022.9699450.

[6] X. Liu *et al.*, "Toward intelligent design and measurement of MEMS acoustic wave resonators," *IEEE Trans. Microw. Theory Techn.*, vol. 74, no. 4, pp. 3478–3490, Apr. 2026, doi: 10.1109/TMTT.2025.3649908.

[7] L. Colombo, M. Ceran, and M. Rinaldi, "Spurious modes metric definition for machine learning aided MEMS design," in *Proc. IEEE Int. Ultrason. Symp. (IUS)*, Venice, Italy, Oct. 2022, pp. 1–3, doi: 10.1109/IUS54386.2022.9957468.

[8] G. Endoh, K. Hashimoto, and M. Yamaguchi, "Surface acoustic wave propagation characterisation by finite-element method and spectral domain analysis," *Jpn. J. Appl. Phys.*, vol. 34, no. 5B, p. 2638, 1995, doi: 10.1143/JJAP.34.2638.